\documentstyle[prd,aps,floats]{revtex}
\begin{document}
\draft
%
%
\input epsf
\renewcommand{\topfraction}{1}
\twocolumn[\hsize\textwidth\columnwidth\hsize\csname
@twocolumnfalse\endcsname
\preprint{EHU-FT/9702, DART-HEP-97/02, SUSSEX-AST 97/2-1, hep-ph/9702368}
\title{Semilocal string formation in two dimensions}
\author{Ana Ach\'{u}carro}
\address{Department of Theoretical Physics, UPV-EHU, Bilbao, Spain\\
Institute for Theoretical Physics, University of Groningen, The 
Netherlands\\
Department of Mathematics, Tufts University, Medford, MA 02155}
\author{Julian Borrill}
\address{Department of Physics and Astronomy, Dartmouth College,
Hanover, NH 03755\\ \cite{jaddress}Center for Particle Astrophysics, 
University of California, Berkeley, CA 94720\\ \cite{jaddress}National Energy 
Research 
Scientific Computing Center, Lawrence Berkeley National Laboratory,\\
University of California, Berkeley, CA 94720}
\author{Andrew R.~Liddle}
\address{Astronomy Centre, University of Sussex, Falmer, Brighton BN1 
9QH, United Kingdom}
\date{\today}
\maketitle
\begin{abstract}
We present a toy model for the investigation of the formation of
semilocal strings, where a planar symmetry is employed to reduce the
system to two dimensions. We approximate the symmetry breaking using an
extension of the Vachaspati--Vilenkin algorithm, where we throw down random 
phases for the scalar fields and then find the gauge field configuration 
which minimizes gradient energy in this fixed scalar background. We show
this procedure reproduces the standard estimate for the formation
rate of cosmic strings.  For semilocal strings the configurations
generated by this method are ambiguous, and we numerically evolve
the configurations forward in time to identify which regions form
strings. We find a significant rate of formation, depending on the ratio of 
couplings $\beta$. For low $\beta$ the formation rate is about one quarter 
that of cosmic strings; this falls as $\beta$ is increased and above $\beta = 
1$, where the string solution is dynamically unstable, no semilocal strings 
form. We show the results are robust by examining 
different initial conditions where, as expected in a thermal environment, the 
initial scalar field need not be on the vacuum. We discuss
 implications for the three-dimensional case.
\end{abstract}
\pacs{PACS numbers: 11.27.+d, 11.15.Ex\\ Preprint: EHU-FT/9702,
DART-HEP-97/02, SUSSEX-AST 97/2-1, hep-ph/9702368}
 
\vskip2pc]

\section{Introduction}

As the universe expands and cools, one expects it to go through a
variety of phase transitions. At each, there is the possibility of the
formation of topological defects~\cite{K76}, the precise types being 
dictated by the physics of the breaking (see Ref.~\cite{reviews} for 
reviews). The standard examples are domain walls, cosmic strings, monopoles 
and texture, where for strings and monopoles one may have either a global or 
a local symmetry breaking. More recently, a new type of defect has been 
introduced which possesses both local and global symmetries, the simplest 
example being the semilocal string~\cite{VA,Mark}.

Of great interest is the number density of defects which form. The
standard technique for making such estimates is a discretization of
space into correlation volumes, which may or may not be accompanied by
a discretization of the vacuum manifold as well. This approach was
pioneered by Vachaspati and Vilenkin \cite{VV} for the case of cosmic
strings, and is easily generalized to monopoles, and, with a bit of
further work, to global textures. These cases provide no particular
difficulty because the presence or absence of defects is dictated
solely by the configuration of the scalar field, regardless of whether
the defects are global or local (though for global strings or
monopoles the defect cores themselves prove not particularly important
for the dynamics of the field ordering).

We wish to consider the formation of semilocal strings. This is
a considerably harder problem, because the existence of string
cores depends crucially on the way in which the scalar and gauge
fields interact. The semilocal string scenario is characterized
by the gauge fields having insufficient degrees of freedom to be
able to completely cancel the scalar field gradients even away
from the core of the string, and the final configuration depends
very much on which of the gradients the gauge field chooses to
cancel. Recently an analysis was made by Nagasawa and Yokoyama
\cite{NY} for a related type of defect, the electroweak string 
\cite{EW} formed at the electroweak phase transition, using
probabilistic arguments applied to thermal configurations. They
concluded a very low production rate, though their analysis may
significantly underestimate the extent to which configurations `near'
string configurations might relax to form proper strings since they
do not consider the effect of the gauge fields in driving the Higgs
values towards zero. Note also that such strings are unstable \cite{EWun}. We 
adopt the alternative strategy of carrying out
numerical simulations of initial configurations in order to determine
those regions which prefer to accumulate magnetic flux to form
strings. In this paper we shall only consider semilocal string
formation in two dimensions, in order to test the method we
propose. We shall turn to the three-dimensional case in a future
publication. We specialize to two dimensions by imposing a planar
symmetry.

We estimate the number density by an extension of the Vachaspati--Vilenkin 
algorithm \cite{VV}. First we
generate an initial configuration by choosing a fixed scalar field
background, upon which we find the gauge field configuration which
minimizes the energy associated with the (covariant) gradients and magnetic 
field energy. Strings are
identified by the location of flux tubes of Nielsen--Olesen type
\cite{NO}.  For cosmic strings we show that this accurately reproduces
the standard result. For semilocal strings, however, the initial
configurations thus generated are ambiguous, featuring a complicated
structure of magnetic flux.  We resolve this ambiguity by numerically
evolving the configuration forward in time, allowing configurations in
the `basin of attraction' of the semilocal string solution to relax
unambiguously into that configuration.

In order to confirm that our results are not an artifact of choosing 
particular initial conditions, we also carry out a set of simulations in 
which the scalar field is not initially on the vacuum manifold. In a truly 
thermal environment representing a second-order phase transition, one expects 
the field to be out of the vacuum as the phase transition completes, while 
the vacuum initial conditions are the expected outcome of a first-order phase 
transition. In any case, we find that the semilocal string density obtained 
in these simulations is extremely similar to that of the vacuum initial 
conditions. 

\section{The Model}

Semilocal strings arise in models with scalar fields and gauge fields,
in which there are insufficient gauge degrees of freedom to completely
cancel off the gradient energy associated with an inhomogeneous scalar
field distribution \cite{VA}. The simplest such theory is one which contains 
two complex scalar fields coupled to a single abelian gauge field. Were
there only one complex scalar field, one would have local cosmic
strings, where the scalar gradients can be completely cancelled by the
gauge field away from the string core. We shall use this standard situation 
as a test of our method.

The main question to be addressed in the formation of semilocal
strings is which of the scalar gradients the gauge field chooses to
cancel. As in any physical system, one expects the driving force will
be the desire of the system to minimize its energy. We therefore
propose the following model for determining conditions after the phase
transition. First, we shall consider the scalar fields alone; when the
symmetry breaks an independent choice of vacuum state is to be made in
each correlation volume, in accordance with the Kibble mechanism
\cite{K76}. At this point, the theory with two complex scalars is
exactly that of a global texture theory. We then fix the scalar
fields, and choose the gauge field to be in the configuration which
minimizes the total energy in this fixed inhomogeneous scalar field
background. 

We work in flat space-time throughout. The Lagrang\-ian for the simplest 
semilocal string model is
\begin{eqnarray}
\label{eL}
{\cal L} & = & \left( \partial_\mu - i A_\mu \right) \phi_1^\dagger \left(
        \partial^\mu + i A^\mu \right) \phi_1 \nonumber \\
 & & + \left( \partial_\mu -
        i A_\mu \right) \phi_2^\dagger \left( \partial^\mu + i A^\mu \right)
        \phi_2 \nonumber \\
 & & - \frac{1}{4} F_{\mu\nu} F^{\mu\nu} - V(\phi_1,\phi_2) \,,
\end{eqnarray}
where $\phi_1$ and $\phi_2$ are two complex scalar fields, $A_\mu$ is
the gauge field, the gauge coupling has been set to unity and
$F_{\mu\nu}$ is the gauge field strength. The potential for the scalar
fields is chosen to be of symmetry breaking type
\begin{equation}
V(\phi_1,\phi_2) = \frac{\beta}{2} \left( |\phi_1|^2 + |\phi_2|^2 - 1 
        \right)^2 \,,
\end{equation}
where the vacuum expectation value of the breaking, like the gauge
coupling, has been scaled to unity as in Ref.~\cite{VA}.  The
remaining parameter $\beta$, which cannot be scaled away, measures the
relative strength of the scalar and gauge coupling constants (more 
specifically, $\beta \equiv m_{{\rm s}}^2/m_{{\rm v}}^2$ is the ratio 
between the squared scalar and vector masses). Its value determines the 
stability of an infinite semilocal string with a Nielsen--Olesen profile: 
the semilocal string is stable for $\beta < 1$, neutrally stable for $\beta 
= 1$ and unstable for $\beta > 1$ \cite{Mark,AKPV}. 

In this paper we shall only consider the two-dimen\-sional situation 
obtained by imposition of a planar symmetry in the `3' direction, $D_3 
\equiv 0$. Working in the gauge $A_3 = 0$, this reduces to the more familiar 
$\partial_3 \equiv 0$. There is a residual gauge freedom which allows the 
further choice $A_0 = 0$. 
It is then clear that the  energy-minimizing configurations will have 
$\partial_0 = 0$, since time derivatives appear quadratically.
 Splitting up the scalar fields into four real 
scalars via $\phi_1 = \psi_1 + i \psi_2$, $\phi_2 = \psi_3 + i \psi_4$, the 
energy density of a static configuration can be written
\begin{eqnarray}
\rho & = & A_1^2+A_2^2 + \frac{1}{2} \left( \partial_1 A_2 - 
        \partial_2 A_1 \right)^2 \nonumber \\ 
 & - & 2 \left[A_1 \psi_1 \partial_1 \psi_2 - A_1 \psi_2 \partial_1 \psi_1
        + A_2 \psi_1 \partial_2 \psi_2 - A_2 \psi_2 \partial_2 \psi_1
        \right. \nonumber \\
 &&   \left. \; + A_1 \psi_3 \partial_1 \psi_4 
        - A_1 \psi_4 \partial_1 \psi_3
        + A_2 \psi_3 \partial_2 \psi_4 - A_2 \psi_4 \partial_2 \psi_3
        \right] \nonumber \\ 
 & + & \left(\partial_1 \psi_1 \right)^2 + \left(\partial_1 \psi_2 
        \right)^2 
        +\left(\partial_1 \psi_3 \right)^2 +\left(\partial_1 \psi_4
        \right)^2 \nonumber \\
 & + & \left(\partial_2 \psi_1 \right)^2 +\left(\partial_2 \psi_2
        \right)^2
        +\left(\partial_2 \psi_3 \right)^2 +\left(\partial_2 \psi_4
        \right)^2 \nonumber \\
 & + & V(\psi_1,\psi_2,\psi_3,\psi_4) \,.
\end{eqnarray}

Having thrown down the scalar field values and fixed them everywhere,
the only terms required to minimize the energy are those depending on
the remaining two components of the gauge field. This contribution to
the energy density can be written as
\begin{equation}
\label{rhoa}
\rho_A = A_1^2 + A_2^2 + 2A_1 f_1 + 2A_2 f_2 + \frac{1}{2} \left[ 
        \partial_1 A_2 - \partial_2 A_1 \right]^2 \!\!,
\end{equation}
where
\begin{equation}
f_i =  -\psi_1 \partial_i \psi_2 + \psi_2 \partial_i \psi_1 -
        \psi_3 \partial_i \psi_4 + \psi_4 \partial_i \psi_3 \,, 
\end{equation}
gives the source terms for the gauge fields, which depend only on the 
(fixed) scalar field configuration. The remaining gauge terms are coupled 
only through the final magnetic flux term. Were it absent, there would be an 
analytic minimization of the energy given by
\begin{equation}
\label{analmin}
A_1({\bf x}) = - f_1({\bf x}) \quad ; \quad A_2({\bf x}) = 
        - f_2({\bf x}) \,.
\end{equation}
This solution provides a good iterative starting point for finding the
true minimum numerically, which will involve trading off some of the flux 
energy in the analytic solution for energy in the other terms. Consequently,
the analytic solution provides an upper limit to the amount of flux
energy possible in the minimum energy configuration.

\section{Initial Conditions and Evolution}

\subsection{Initial conditions}

There is no hope of an analytic minimization of Eq.~(\ref{rhoa}) for a
general scalar field configuration, so the problem must be dealt with
via discretization. The problem being two-dimensional, we break up the
space into a grid of dimension $N \times N$. The first stage is to lay
down random phases for the scalar field to form the background for the
energy minimization. The scalar field problem is just that of a global
texture theory, which was examined in Ref.~\cite{BCL1}. However, in
that paper the vacuum manifold was discretized, which is unnecessary
(and even disadvantageous) here; instead random locations on the
vacuum manifold are chosen. Because the manifold has three-sphere
topology, the scalar field can be in vacuum everywhere, in contrast to
the usual cases where topology may force the scalar field to leave the
vacuum manifold.  However, once the gauge fields are introduced, they may
make it energetically favourable for the scalar field to move away from
the vacuum manifold, for example by allowing the magnetic flux to
concentrate in areas of restored symmetry, and in such locations a
semilocal string will have formed.

In order that the fields can be treated as fairly continuous on the
grid, one should not throw down a completely random value at each
grid-point.  Instead, we choose the correlation length to be some
number of grid-points and throw down phases on that coarser grid. The
correlation length is chosen to be $2^p$ grid lengths, where $p$ is an
integer. The field is then interpolated recursively onto the full grid
by bisection. Thus our two-dimensional square grid with $2^m$
grid-points on a side contains $2^{2(m-p)}$ initial correlation
volumes. Such a procedure was used in Ref.~\cite{BCL3} to set up
initial conditions for dynamical texture evolution. The simplest
interpolation is somewhat discontinuous at the boundaries of the
initial grid, which did not matter in Ref.~\cite{BCL3} as they were
quickly smoothed by dynamical evolution. Here we use a cubic
interpolation in order to provide a smoother configuration that allow
examination of the initial configuration itself. The gauge fields are
subsequently chosen to minimize (covariant) gradients.  We
impose periodic boundary conditions; these enforce zero net flux
through the grid, but this turns out to be unimportant provided the
grid is sufficiently large.

We minimize the energy via a Monte Carlo procedure, starting with the 
analytic minimum Eq.~(\ref{analmin}), by choosing a grid-point and updating 
the gauge field values to new, energy-minimizing, values. After each point 
has been visited many times, the total energy is found to stabilize --- we 
experimented with variation of the random sequence of points visited and 
also the initial conditions to confirm that the system did not get trapped 
in a false minimum.

This procedure is sufficient to obtain results for cosmic strings, reported 
in the next Section, simply by identifying visually (or automatically via a 
numerical analysis routine) the location of flux 
tubes in the configuration. For semilocal strings it proves impossible to 
decide from the iterated configuration which regions will become semilocal 
strings, as a very complex flux structure results due to the more 
complicated scalar field background. In order to resolve this ambiguity, it 
proves necessary to evolve the initial configuration forward in time to see 
what happens. It is also clear that this is necessary to distinguish the 
situations of stable and unstable semilocal strings, determined by the 
choice of $\beta$. The procedure thus far has been independent of $\beta$, 
but for $\beta > 1$ we expect the flux to dissipate, while for $\beta < 1$ 
some stable strings might form.

\subsection{Evolution}

Both analytical \cite{P,H,BB} and numerical \cite{AKPV} methods have
already been used to study issues concerned with stability of the
string configurations for general $\beta$, and the cosmological 
implications, but have not been used to examine the evolution of initial 
configurations such as we have generated here.

With our gauge choices and with planar symmetry, the equations of motion 
from the Lagrangian of Eq.~(\ref{eL}) are
\begin{eqnarray}
\ddot{\psi}_i - \nabla^2 \psi_i + \beta \left( \psi_1^2 +
        \psi_2^2 + \psi_3^2 + \psi_4^2 - 1 \right) \psi_i \\
 \hspace*{1cm} + A^2 \psi_i + (-1)^{j} \left( 2 A.\nabla + \nabla.A \right) 
        \psi_j = 0  \,, \nonumber
\end{eqnarray}
(where $j$ is the complement of $i$ --- $1 \leftrightarrow 2$, $3
\leftrightarrow 4$, and dots are of course time derivatives) for the 
scalar fields and
\begin{eqnarray}
\ddot{A}_i - \nabla^2 A_i + \partial_i \nabla.A + 2 \left(
        \psi_1 \stackrel{\leftrightarrow}{\partial_i} \psi_2 +
        \psi_3 \stackrel{\leftrightarrow}{\partial_i} \psi_4 \right) \\
 + 2 A_i \left( \psi_1^2 + \psi_2^2 + \psi_3^2 + \psi_4^2 \right)
        = 0 \,, \nonumber
\end{eqnarray}
for the gauge fields, together with the constraint (from the gauge choice)
\begin{equation}
2 \left( \psi_1 \stackrel{\leftrightarrow}{\partial_0} \psi_2 +
        \psi_3 \stackrel{\leftrightarrow}{\partial_0} \psi_4 \right)+
        \partial_i \dot{A}_i = 0 \,,
\end{equation}
which is used to test the stability of the code. The arrows indicate 
asymmetric derivatives.

This system is discretized using a standard staggered leapfrog
method; however, to reduce its relaxation time we also add an
{\it ad hoc} dissipation term to each equation --- $\eta
\dot{\psi}_i$ and $\eta \dot{A}_i$ respectively. In an expanding
Universe the expansion rate would play such a role, though $\eta$ would 
typically not be constant. We tested a wide range of strengths of 
dissipation, and checked that it did not affect the end result, only the 
time the system took to adopt a distinctive structure. The simulations we 
display later used $\eta = 0.5$. Fortunately we expect the long-range 
interactions between static parallel semilocal strings to be negligible 
\cite{AKPV} so we can take the number of flux tubes remaining at the end of 
the evolution to be a good estimate of the initial number of strings (and in 
any case a lower limit). This expectation is also 
borne out by examination of the flux tubes' evolution over time.

\section{Results}

\subsection{Cosmic strings}

To check our general strategy, we first test the performance
of our codes for the case of cosmic strings. This simply involves
ignoring one of the (complex) scalar fields, setting $\psi_3 =
\psi_4 = 0$. This makes the defects topological, and the flux
tubes formed now map out the locations of winding in the scalar
field configurations.

\begin{figure*}
\centering 
\leavevmode\epsfysize=10.9cm \epsfbox{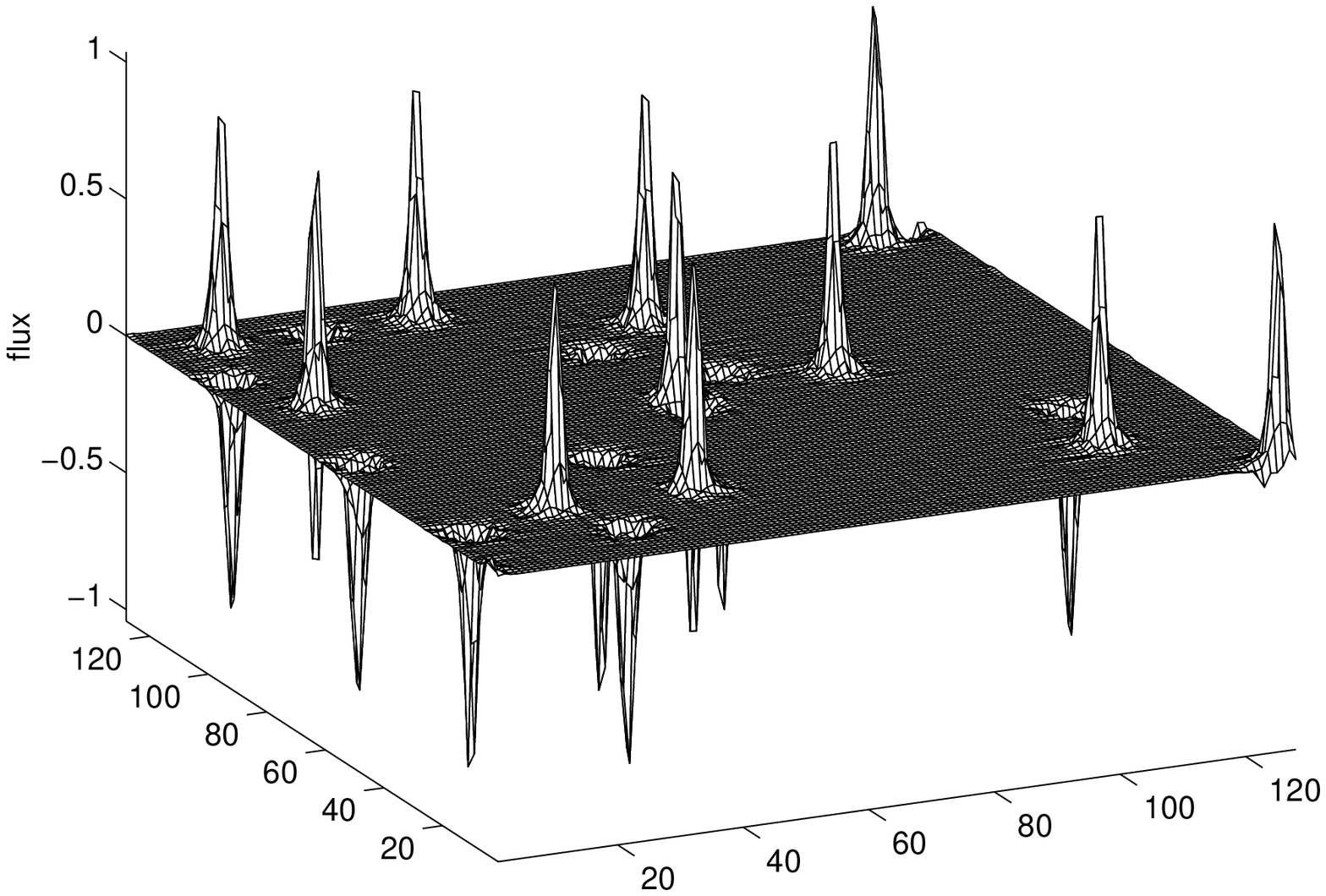}\\
\leavevmode\epsfysize=10.9cm \epsfbox{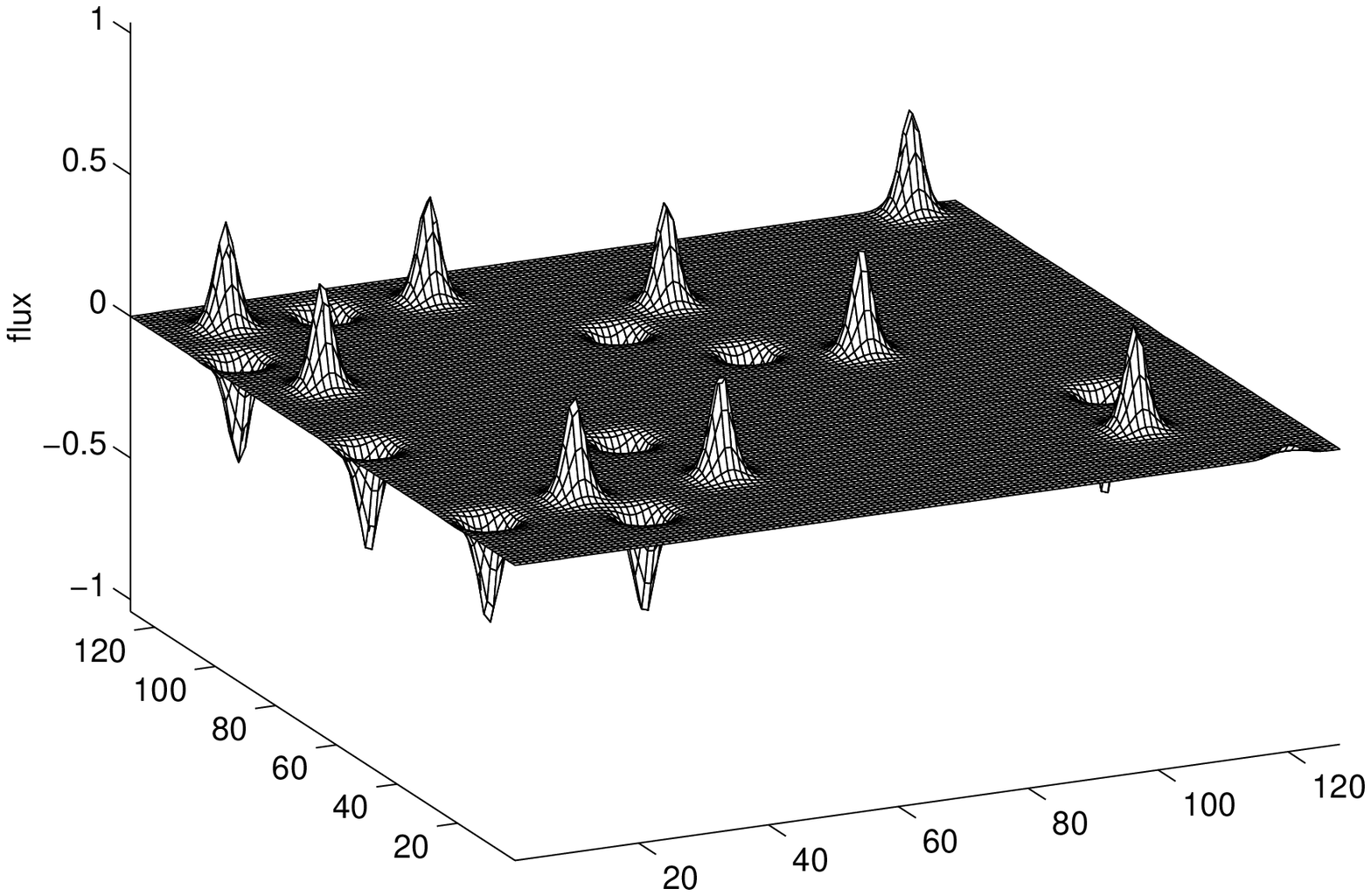}\\
\caption[semi1]{\label{semi1} The flux tube structure of a cosmic string 
network. The flux tubes tend to line up on the sub-grid formed by the
initial correlation length of 16 gridpoints. The spikes in the upper
panel are taller and narrower than the corresponding Nielsen--Olesen 
vortices which result after evolution forward in time with $\beta = 0.05$, 
as shown in the lower panel at $t=100$. Two flux pairs annihilate during this 
evolution.}
\end{figure*} 

In this case the flux tube structure is very well defined after
the iteration has been used to produce the initial conditions, and we can 
directly estimate the number
density of strings formed without needing any dynamical
evolution. Using a $128^2$ grid and smoothing over every 16
grid-points we have 64 initial correlation volumes. An iterated 
configuration, with 22 flux tubes, is shown in the upper 
panel of Fig.~1. Examining 10 different initial field 
configurations we obtain a cosmic string number density
\begin{equation}
n_{{\rm cs}} = 0.32 \pm 0.02 
\end{equation}
per correlation volume. The error quoted is the standard 
error on the mean, but note that the uncertainty from discretization is somewhat 
larger.

This result agrees perfectly with the standard Vachaspati--Vilenkin
algorithm on a 2-dimensional square lattice with a continuous field.
Assuming the geodesic rule (i.e.~that the field always covers the
shortest path on the vacuum manifold along a grid edge), then the
probability of string formation can be computed analytically. For
example, take one corner to be at zero phase, take the diagonally
opposite corner to be uniformly distributed in phase and compute the
probability that the remaining two corners have phases in the right
place to give a winding. The probability is given by
\begin{equation}
n_{{\rm cs}}^{{\rm VV}} = \frac{2}{(2\pi)^3} \int_0^{2\pi} 
        (2\pi - \theta) \, \theta \, d\theta = \frac{1}{3} \,,
\end{equation}
in agreement with what we found above. The good agreement arises because the 
simulations are themselves carried out on a square lattice, and confirms that 
our identification of cosmic 
strings via the flux tube structure works extremely well. The greatest 
uncertainty in the estimate comes from the choice of 
discretization; for example had we instead used a triangular lattice the 
Vachaspati--Vilenkin algorithm would give a formation rate of 
$1/4$ per triangle rather than $1/3$ per square \cite{Prok}, though one 
would also have to account for the difference in area.

Note that although the flux tubes in the upper panel of Fig.~\ref{semi1} are 
very clearly defined, they are not
Nielsen--Olesen vortices \cite{NO}, because the scalar field is
everywhere in vacuum. Their width is governed by the inverse vector
mass. Once the scalar fields are allowed to relax, the flux tubes become 
Nielsen--Olesen vortices. The result of dynamical evolution, for one hundred 
time units, is shown in the 
lower panel of Fig.~1, with $\beta$ chosen as 0.05. With that value, the 
inverse scalar mass is larger than the inverse vector mass (recall that 
$\beta \equiv m_{{\rm s}}^2/m_{{\rm v}}^2$) and so the total flux, while 
remaining $2 \pi$, spreads over a larger area.
During the evolution between the upper and lower panels, two pairs of flux 
tubes annihilate leaving 18 out of the original 22. Such annihilations are 
expected as the fields evolve and develop longer-range correlations, though 
this is inhibited by our artificial viscosity. 

\begin{figure*}
\centering 
\leavevmode\epsfysize=10.9cm \epsfbox{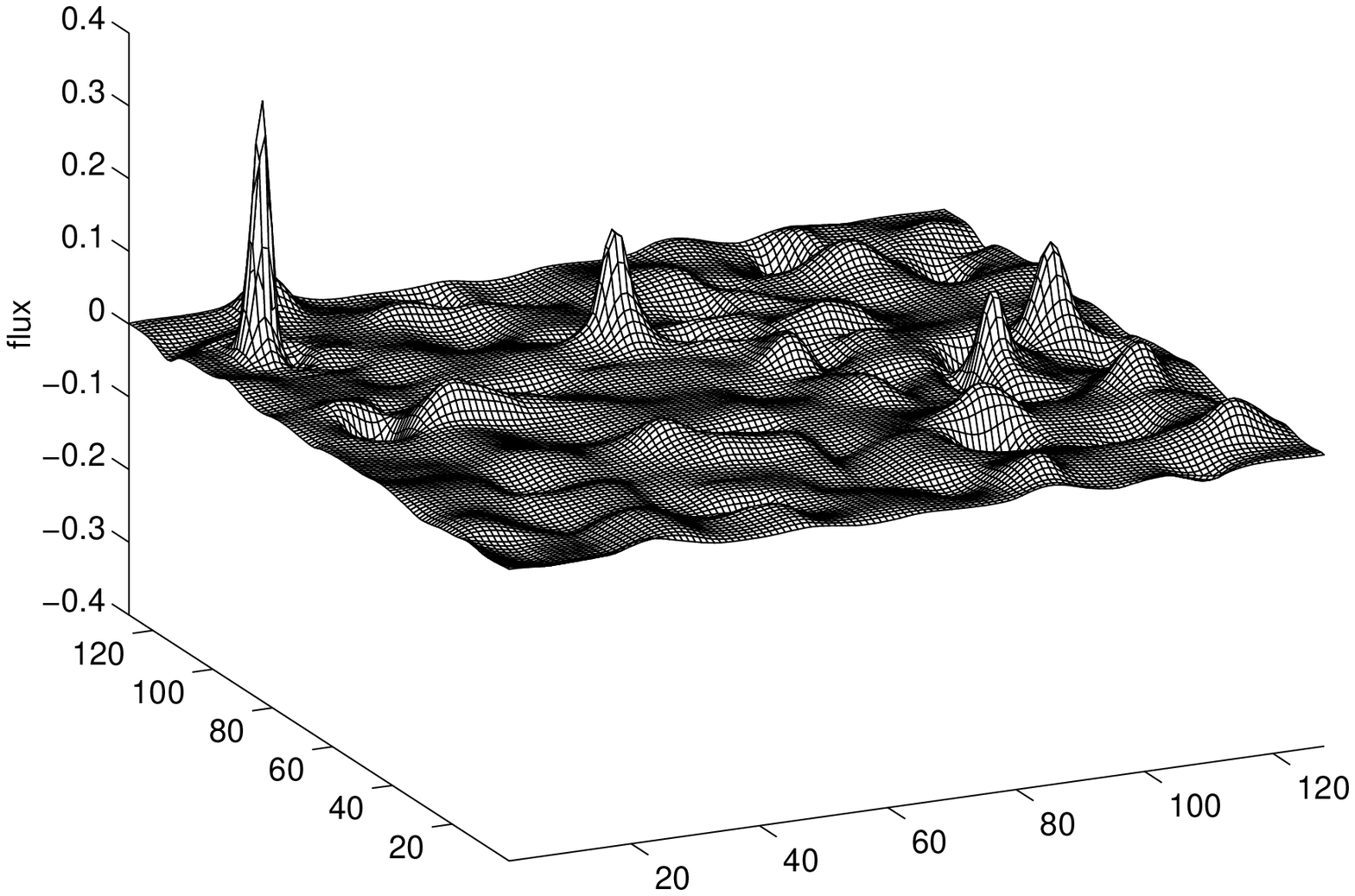}\\ 
\leavevmode\epsfysize=10.9cm \epsfbox{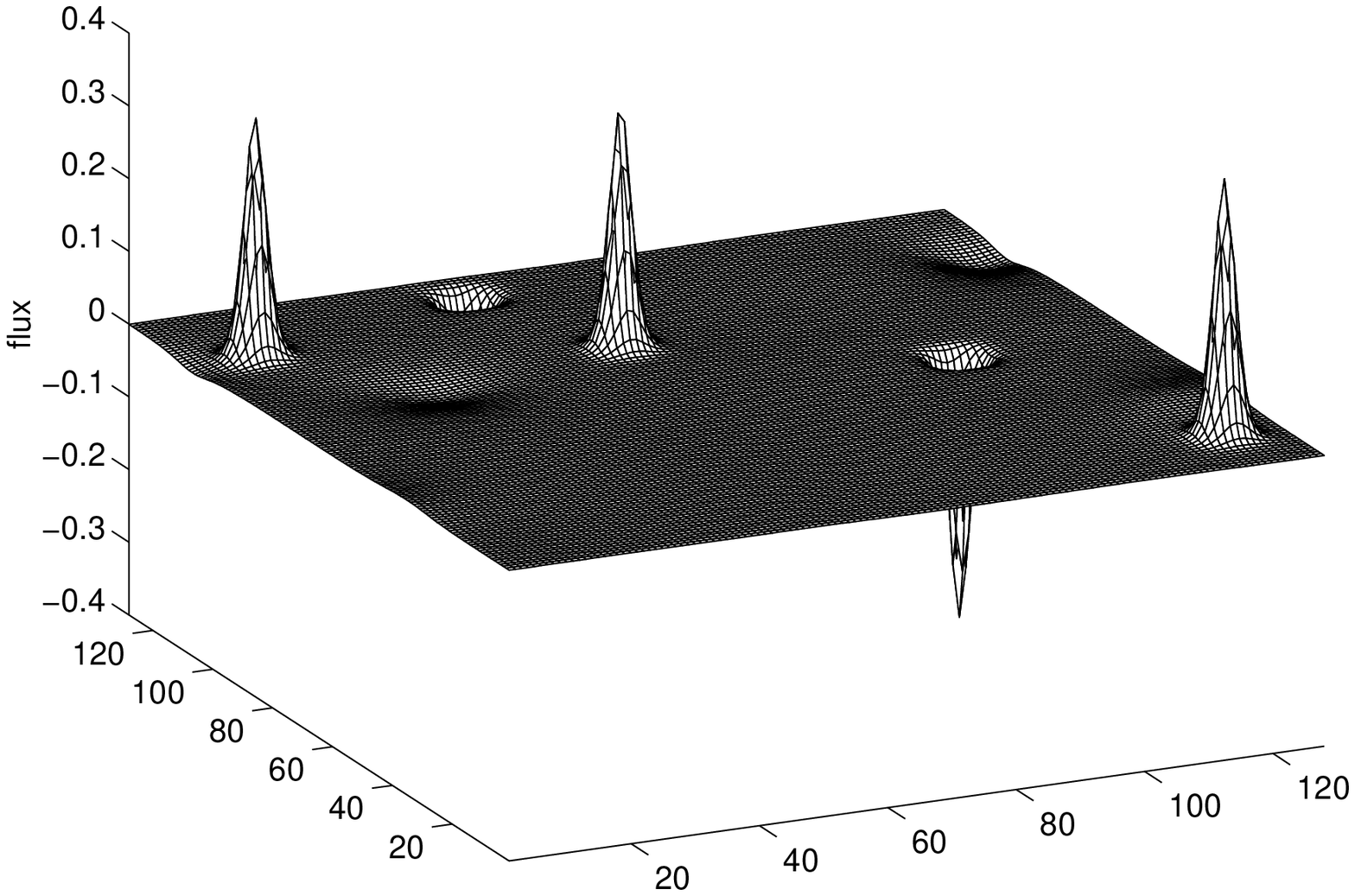}\\ 
\caption[semi2]{\label{semi2} The flux tube structure in a semilocal string 
  simulation. The upper panel ($t=0$) shows the initial condition
  after iteration, which is much more complicated than in the cosmic
  string case and not readily interpreted. The lower panel shows the
  configuration resolved into five flux tubes by a short period of
  dynamical evolution ($t=100$). Note that the vertical scale differs
  from that of Fig.~1; the final flux tubes have the same profile as in that
  case.}
\end{figure*} 

\subsection{Semilocal strings}

For semilocal strings, the initial configurations generated by iteration 
prove much more ambiguous, because there is no topological constraint. One 
sees a complicated flux structure with extrema of different heights, and it 
is far from clear which of these, if any, might evolve to form semilocal 
strings; an example is shown in the top panel of Fig.~2. In any
case, something has to be done to account for the influence of the parameter 
$\beta$, which determines the stability of the string configurations. We do 
this by evolving the system forward in time, including a dissipation term as 
described above to aid the relaxation.\footnote{A 100 frame {\tt mpeg} movie 
(0.5Mb) of the entire simulation can be viewed at {\tt 
http://star-www.cpes.susx.ac.uk/\\people/arl\underline{~}recent.html} and we 
shall maintain it there for as long as we can.} In fact, carrying out this 
dynamical evolution renders the energy-minimization process redundant,
since the early stages of evolution carry out this role anyway. 


As anticipated, in the unstable regime $\beta > 1$ the
flux quickly dissipates leaving no strings. In the
stable regime $\beta < 1$ stringlike features persist.
The lower panel of Fig.~2 shows the result of evolving the
configuration of the upper panel forward in time for one hundred time
units (which is about a factor of two longer than the timescale found
in Ref.~\cite{AKPV} for the relaxation into semilocal strings of
nearby configurations with non-zero winding number).  Comparing the two 
panels indicates that some, but not all, of the more prominent flux features
in the initial conditions do indeed relax into semilocal strings, that
is, Nielsen--Olesen vortices identical to those shown in the bottom
panel of Fig.~\ref{semi1} for the cosmic string case (note the
different vertical scales of Figs.~1 and 2).

In order to automate the counting of flux tubes for the large number of 
simulations we have obtained, we adopt a criterion that any isolated extrenum 
in the magnetic flux which is at least half that of a perfect 
Neilsen--Olesen vortex is identified as a string. We tested this criterion on 
selected simulations which we studied in detail ourselves, and concluded it 
gives an accurate counting. This enabled us to carry out a large number of 
runs to bring down the statistical errors on our results.

For each of seven different values of $\beta$, we take 10 initial configurations 
on a $128^2$ grid smoothed over every $16$ grid-points. 
For $\beta < 1$ we find a formation rate which depends on $\beta$, and which 
is always significant, though not as large as the number density found in the 
cosmic string case. At the smallest $\beta$ we were able to simulate, 0.05, 
the formation rate is about one quarter of that of cosmic strings. (If 
$\beta$ is lowered further, the scalar string cores become too wide to fit 
into a correlation volume.) Note that the 
lower number density means that annihilations during the early stages of 
evolution are less likely than in the cosmic string case, so would lead to 
an even smaller correction here.

\begin{figure}
\centering 
\leavevmode\epsfysize=6cm \epsfbox{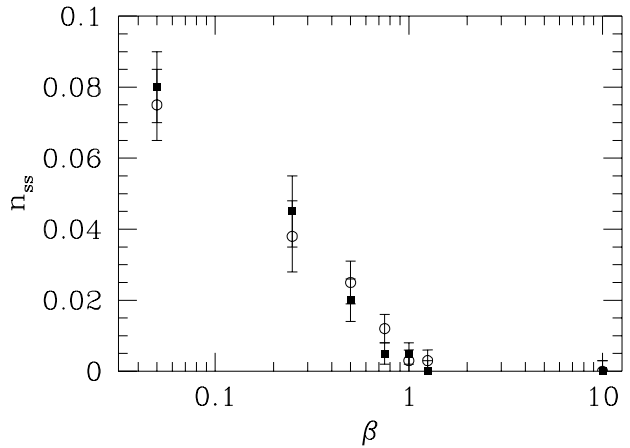}\\ 
\caption[semi3]{\label{semi3} The number of semilocal strings formed per 
initial correlation volume. Each point is an average over ten simulations. 
Squares indicate that the vacuum initial conditions of Section IV were used, 
while open circles indicate that the non-vacuum initial conditions described 
in Section V were used.}
\end{figure} 

Support for our results comes from an analysis \cite{AKPV} in which simple 
configurations were permitted to relax into
semilocal strings. It was found that configurations some way
from a maximal circle (about one quarter of the way from the great
circle for low values of $\beta$, decreasing to zero as $\beta$ approaches one) 
would relax into semilocal strings if
$\beta < 1$. Consider the values of the Higgs in four points on the
lattice defining a grid square. All values lie on the three sphere.
If we project all four points onto the most favourable maximal circle,
then the probability of winding is identical to the
Vachaspati--Vilenkin calculation, i.e.~1/3. Since the points are away
from the maximal circle the suppression to the probability that they
will still form a semilocal string can be estimated by whether the
point which is furthest from the maximal circle is closer than a
certain cutoff distance $D_{{\rm c}}$, and will be proportional to the
volume of a fringe of width $2D_{{\rm c}}$ around the equator (over
the total volume of the three sphere). Ref.~\cite{AKPV} gives an
estimate of $D_{{\rm c}}$, and the probability obtained is consistent
with Fig.~3. 

Notice that the final configuration illustrated in Fig.~\ref{semi2}
does not have the same number of flux tubes of upward and downward
pointing flux, despite the zero total flux boundary condition. The
extra flux resides in small `nodules' of flux that can be seen in the
figure; although the maximum flux density is tiny, they are nevertheless 
quite
diffuse and contain a significant amount of total flux.  The structure
of these nodules changes if the artificial viscosity is varied,
indicating that they are in fact an artifact made long-lived by the
viscous term. However, we note that if expansion was taken into
account it would have some properties similar to our viscosity, and may
permit such `skyrmionic' configurations to be long-lived. Such a suggestion 
has been made by Benson and Bucher~\cite{BB}.

\section{Relaxing the initial conditions}

In order to confirm the robustness of our conclusions, we also carried out 
the same number of simulations with different initial conditions. For a 
second-order 
phase transition, one expects the scalar field to be out of the vacuum as the 
transition takes place. This has been studied by Ye and 
Brandenberger~\cite{YB}, and we closely follow their strategy. They chose the 
scalar field magnitude randomly from a uniform distribution; we adopt a 
similar approach though using a gaussian distribution instead. We examined 
different 
choices for the width of the gaussian, and any reasonable choice makes no 
difference to the results. We shall display those for when the gaussian has a 
dispersion equal to the vacuum expectation value, which is the natural value 
at the phase transition.

We find startlingly good agreement with the vacuum initial conditions, as 
shown in Fig.~3. The same general shape for the $\beta$-dependence is found, 
and the number of strings identified per simulation is extremely close, with 
the error bars overlapping. For cosmic strings we found that the agreement 
between the two 
sets of simulations is not quite as good, though still impressive; some 
differences are expected as the initial correlation length will have 
changed, and we conclude that these changes are less significant for 
semilocal strings than cosmic strings.

\section{Conclusions}

In this paper we have proposed a method of investigating semilocal
string formation by following the magnetic flux, and tested it by
restricting the problem to two dimensions by imposing a planar
symmetry. We have shown that our method can accurately reproduce the
standard result for cosmic strings.  We find a very significant rate
of semilocal string formation; about one quarter the rate of cosmic string 
formation for the lowest $\beta$ we tested, falling towards zero as the 
unstable regime is reached at $\beta = 1$. This result is robust to significant 
changes in the initial conditions.

Although superficially our result for the semilocal string density appears 
to be in disagreement with the very low probability of formation found by 
Nagasawa and Yokoyama for electroweak strings \cite{NY}, a direct comparison 
cannot be made at this point. Their computation concerns the initial 
configuration only. It does not take into account the effect of the 
$A^2\phi^2$ term in the Lagrangian (which is usually responsible for the 
Meissner effect in superconductors), which will tend to drive the value of 
the Higgs field towards zero in regions where there is a high concentration 
of magnetic flux, in competition with the potential term which favours
the vacuum value of the Higgs. Our simulations seem to indicate that
the backreaction of the gauge fields is substantial, at least in the
case of semilocal strings, and we expect this to be true also for 
electroweak strings as long as the SU(2) coupling is sufficiently small to 
allow stable strings. 
This may lead to string formation during the subsequent evolution. Further, 
the electroweak strings they consider are in fact unstable \cite{EWun}, and 
were we to carry out simulations of that case we would conclude a zero 
formation rate.\footnote{We also note that in the case of a first-order phase 
transition the use of the geodesic rule has been questioned by  Saffin and 
Copeland~\cite{SC}.}

However, our simulations are only two-dimensional, and the three-dimensional 
implications are not straightforward. For cosmic strings, the 
two-dimensional formation rate can be used more or less directly to estimate 
the three-dimensional formation rate, because one can rely on the topology 
to ensure that the strings cannot end. Semilocal strings need not be 
infinite; there is the possibility that the strings might end in a 
diffuse cloud of flux resembling a global monopole. Further, the analytic 
infinite semilocal string 
solution corresponds to a winding of the gauge fields round a fixed great 
circle of the $S^3$ vacuum manifold, whereas in a realistic 
situation even if the winding condition continues to be obeyed as one moves 
along a putative string, the winding may correspond to different great 
circles at different points along the string and whether or not such a 
situation can be dynamically stable remains unexplored. Consequently, as yet 
we have no guidance as to how easily a configuration may find itself within 
a `basin of attraction' of a semilocal string configuration. A complete 
picture of the formation rate must await full three-dimensional simulations, 
which are currently underway.

\section*{Acknowledgments}

AA is partially supported by NSF grant PHY-9309364, CICYT grant
AEN96-1668 and UPV grant 063.310-EB225/95. JB is supported by NSF grant 
PHY-9453431 and ARL by the Royal Society. This work was also
supported in part by the European Commission under the Human Capital
and Mobility programme, contract number CHRX-CT94-0423, and by a NATO 
Collaborative Research Grant `Cosmological Phase Transitions' no. SA-5-2-05 
(CRG 930904) 1082/93/JARC-501. ARL is grateful to the Kapteyn Institute and 
the University of Groningen for their hospitality, and AA and
ARL thank the Isaac Newton Institute where this work was begun.
We thank Mark Hindmarsh, Konrad Kuijken and Tanmay Vachaspati
for many useful discussions.


\end{document}